\documentclass[sigconf]{acmart}

\usepackage{subcaption}
\usepackage{graphicx}
\usepackage{multirow}
\usepackage{booktabs} % For formal tables
\usepackage{amssymb}% http://ctan.org/pkg/amssymb
\usepackage{pifont}% http://ctan.org/pkg/pifont
\usepackage{listings}
\usepackage{balance}
\usepackage{color}
\usepackage[normalem]{ulem}
\usepackage{url}
\usepackage{hyperref}

\usepackage{courier}
\newcommand{\cmark}{\ding{51}}%
\newcommand{\xmark}{\ding{55}}%
\newcommand{\para}[1]{{\vspace{4pt} \bf \noindent #1 \hspace{10pt}}}

\setcopyright{none}

\renewcommand\footnotetextcopyrightpermission[1]{} % removes footnote with conference information in first column
\begin{document}
\author{Hang Hu}
\affiliation{
\institution{Virginia Tech}}
\email{hanghu@vt.edu}

\author{Limin Yang}
\affiliation{
\institution{University of Illinois Urbana-Champaign}}
\email{liminy2@illinois.edu}

\author{Shihan Lin}
\affiliation{
\institution{Fudan University}}
\email{shlin15@fudan.edu.cn}

\author{Gang Wang}
\affiliation{
\institution{University of Illinois Urbana-Champaign}}
\email{gangw@illinois.edu}

\begin{abstract}
The popularity of smart-home assistant systems such as Amazon Alexa and Google Home leads to a booming third-party application market (over 70,000 applications across the two stores). While existing works have revealed security issues in these systems, it is not well understood how to help application developers to enforce security requirements. In this paper, we perform a preliminary case study to examine the security vetting mechanisms adopted by Amazon Alexa and Google Home app stores. 
With a focus on the authentication mechanisms between Alexa/Google cloud and third-party application servers ({\em i.e.} end-points), we show the current security vetting is insufficient as developer mistakes cannot be effectively detected and notified. 
A weak authentication would allow attackers to spoof the cloud to insert/retrieve data into/from the application endpoints. We validate the attack through ethical proof-of-concept experiments. To confirm vulnerable applications have indeed passed the security vetting and entered the markets, we develop a heuristic-based searching method. We find 219 real-world Alexa endpoints that carry the vulnerability, many of which are related to critical applications that control smart home devices and electronic cars. We have notified Amazon and Google about our findings and offered our suggestions to mitigate the issue.

\end{abstract}

% \title{Peeking Over the Edge: Spoofing the Cloud Service of Smart Home Assistant Systems}

%%% suggested new title %%%%
\title{Security Vetting Process of Smart-home Assistant Applications: A First Look and Case Studies}

\maketitle

\acmConference[]{The Web Conference}{April 2020}{Taibei}
% \acmYear{2020}
% \copyrightyear{2019}
\pagestyle{plain}

%%% gang's new logic flow %%%%
% - Smart-home assistant systems have a new app markets
% - Security is important
% - Existing works have looked at security vulnerabilities in these devices
% - But did not see how to help developers mitigate risks
% - Unlike mobile apps, developers are less familiar
% - In this paper, we perform a case study on X and Y, with a focus on authentication mechanisms. 

\section{introduction}
\label{sec:intro}
Smart home assistant systems such as Amazon Alexa and Google Home are entering tens of millions of households today~\cite{forbesiot18}. As a result, the corresponding app marketplace is also expanding quickly. Just like installing apps on smartphones, users can enable third-party applications for smart-assistant devices. These applications are called ``skills'' or ``actions''. So far there are collectively more than 70,000 skills available~\cite{amazonskill2019, googleskill2019}, many of which are security/safety-critical. For example, there are skills that allow users to manage bank accounts, place shopping orders, and control smart-home devices through a voice interface. 

Considering the sensitive nature of smart-home assistants, researchers have looked into the security aspects of these systems and their third-party applications. For example, recent studies show that it is possible to craft a voice clip with hidden commands embedded that are recognizable by the Alexa device but not by human observers~\cite{carlini2016hidden, zhang2017dolphinattack,ndss2019,commandsoung}. In addition, researchers demonstrate the feasibility of a ``skill squatting'' attack to invoke a malicious application whose name sounds like the legitimate one~\cite{kumar2018skill, zhangdangerous}. A recent survey study~\cite{alrawisok} investigated the network interfaces of many IoT devices (including smart-assistant devices) to reveal their weak encryptions and unpatched OS/software. While most existing studies focus on the system and device-level flaws, limited efforts are investigated to {\em vetting the security of third-party applications}, and more importantly, {\em helping developers to improve the security of their applications}.

In this paper, we perform a preliminary {\em case study} to examine the mechanisms that Amazon and Google implemented to vet the security of third-party applications for their smart home assistants. More specifically, before a third-party application (or ``skill'') can be published to the app stores, they must go through a series of automated tests and manual vetting. In this paper, we seek to understand (1) what aspects the security vetting process is focused on, and (2) how effective the vetting process is to help developers to improve security. 

As a preliminary study, we focus on the authentication mechanism used by the third-party application's server (called ``endpoint'') to authenticate the Alexa/Google cloud (namely, cloud authentication). We choose cloud authentication because cloud-endpoint interaction is a key component that makes smart-home assistant skills\footnote{For convenience, we refer smart-home assistant application as ``skills'' for both Amazon Alexa and Google Home.} structurally different from the conventional smartphone apps. Smart-home assistant skills need to route their traffic to a central cloud to translate a voice command to an API call in order to interact with the application server.

\para{Method and Key Observations.} Amazon Alexa runs both {\em automated vetting} and {\em manual vetting} before a skill can be published, while Google Home only runs {\em manual vetting}. 
Our methodology is to build our own (vulnerable) skills and walk them through the required testing to understand the vetting process. 
Our results show concerning issues in terms of the enforcement of cloud authentication. 
First, we find that the Google Home vetting process does not require the endpoints to authenticate the cloud and their queries, which leaves the endpoints vulnerable to spoofed queries. 
Second, Amazon Alexa requires skills to perform cloud authentication, but does a poor job enforcing it on third-party developers. 
Alexa performs automated vetting that is supposed to detect and inform developer mistakes in the skill implementation. However, the security tests are erroneous and have missed important checks ({\em e.g.}, application identifiers). 
As a result, a vulnerable skill, in theory, can pass the security vetting process to enter the app store.

\para{Proof-of-Concept.} To illustrate the problem, we run controlled experiments to show how an outsider can spoof the cloud to query the target endpoint. More specifically, an attacker can build its own skill application, and use this skill to obtain a valid signature from the cloud for the attack traffic. Then the attacker can replay the signed traffic to attack the target endpoints. The attack is possible because the cloud uses the same private key to sign all the traffic for all the skills. The signature obtained by the attacker's skill works on the victim endpoint too. We validate the feasibility of the attack and show that vulnerable skills can bypass both the automated tests and the manual vetting process to enter the app markets.

\para{Vulnerable Applications in the Wild.} 
To confirm that there are indeed vulnerable skills in practice, we perform a scanning experiment. Since all Google Home skills are by default vulnerable, this experiment focused on searching for vulnerable Alexa skills. 
We leverage ZMap to locate live HTTPS hosts and replay a spoofed but {\em non-intrusive} query to see if a given HTTPS host returns a valid response. 
In this way, we located 219 vulnerable real-world Alexa endpoints. 
A closer analysis shows that some of these vulnerable endpoints are related to important skills such as those that control electric cars, smart locks, thermostats, security cameras, and watering systems.

We make three main contributions:
\begin{itemize}

\item 
First, we present the first empirical analysis of the security vetting process used by Amazon and Google to vet their smart-home assistant skills. We find that the current vetting process is insufficient to identify and notify developers of the authentication issues in their endpoints.

\item Second, we validate the security problem by running a proof-of-concept cloud spoofing attack, in an {\em ethical manner}.   

\item Third, we discover real-world applications that carry the vulnerability. We notified Amazon and Google about our findings and offered our suggestions to mitigate the issue.  
\end{itemize}

Our work makes a concrete step towards enhancing the security vetting process of smart-assistant applications. Give the increasing number and diversity of IoT applications and web of things, we argue that developers need more help to build their applications in a secure way. Effective security vetting and feedback will be critical to improving the security of the IoT ecosystem.

\section{Background \& Motivation}
% There are about 50 million households that have a smart assistant at home. 

%and 11.2\% of the smart speaker market.
% According to another report published at the same year~\cite{google_beats_alexa},
% Google and Amazon sold 3.2 million and 2.5 million units of their smart speakers during the first quater of 2018.

% \begin{figure}[t]
% \centering
% \includegraphics[width=0.32\textwidth]{plot/amex.eps}
% \caption{
% The Alexa skill of American Express.
% }
% % \vspace{-0.1in}
% \label{fig:amex}
% \end{figure}
% Figure~\ref{fig:amex} shows the webpage of the American Express skill. 
% People can use this skill to manage their bank account and even transfer money through the smart assistant. The webpage lists the skill's basic functions, supported command-lines, developer information, and user reviews. Later in \S\ref{sec:dataset}, we crawled both skill stores, and found 32,289 Alexa skills and 2,237 Google actions, as of June 2018.

We start by introducing the background for third-party applications in Amazon Alexa and Google Home. We then discuss the security vetting process on both platforms. 

\begin{figure}[t]
\centering
\includegraphics[width=0.33\textwidth]{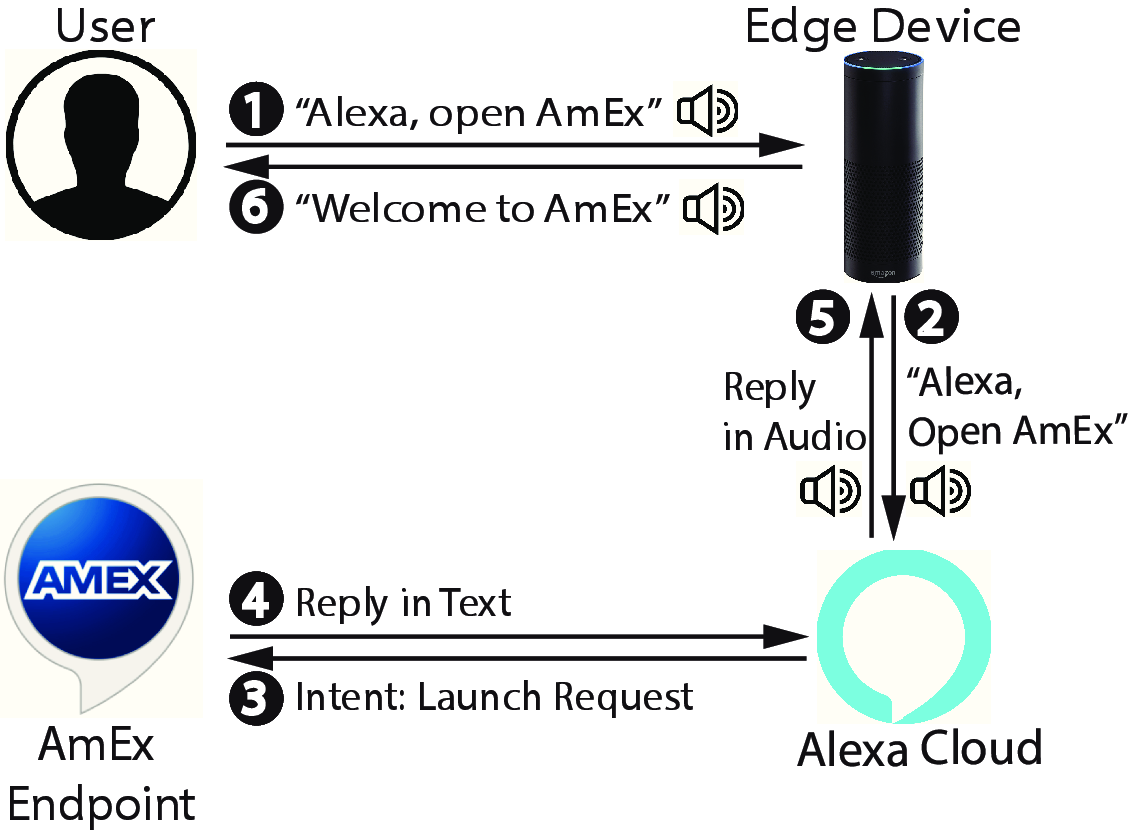}
\caption{The execution of a simple voice command.
}
\vspace{-0.2in}
\label{fig:alexa}
\end{figure}

\para{Alexa and Google Home Skills.} Both platforms support third-party applications, which are called ``Skills'' on Alexa and are called ``Actions'' on Google Home. We use ``skill'' for convenience. They work in a similar way. Figure~\ref{fig:alexa} shows how a user interacts with a skill.  (\ding{182}) a user talks to the edge device to issue a voice command. (\ding{183}) the edge device passes the audio to the Alexa cloud. (\ding{184}) the cloud is responsible to convert the speech to text, and recognize which skill the user is trying to interact with. In addition, the cloud infers the ``intent'' of the command and match it with the known intents pre-defined by the skill developers. Here, {\em intent} is a short string to represent a functionality of the skill. After that, the cloud sends an HTTPS request to the skill's endpoint ({\em i.e.}, a web server). (\ding{185}) the endpoint sends the response back, and (\ding{186}) the cloud converts the text-based response to audio, and (\ding{187}) plays it at the edge-device. Note that the edge device never directly interact with the endpoint, and every request is routed through the cloud. For certain skills, users need to explicitly ``enable'' them in the skill store. However, many skills can be directly triggered/used by calling the skill's name. 

Skill developers need to implement the endpoint to respond to user requests. For simple skills that do not require a database, both Alexa and Google provide a ``serverless'' option for developers to hard-code the responses in the cloud. For sophisticated skills, an endpoint is needed. 

\para{Authentication between Entities.} The system contains three main entities: the edge device, the cloud, and the endpoint. Both Alexa and Google require HTTPS for all network communications. This also helps the ``clients'' to authenticate the ``servers'': The edge device can verify the identity of the cloud, and the cloud can also verify the identity of the endpoint. 

% \begin{table}[t]
% \centering
% \resizebox{0.49\textwidth}{!}{
% \begin{tabular}{|c|c|c|c|c|c|}
% \hline
% System & \begin{tabular}[c]{@{}l@{}}HTTPS\\ Endpoint\end{tabular} &  \begin{tabular}[c]{@{}c@{}}Cloud Auth.\\ \end{tabular}& \begin{tabular}[c]{@{}c@{}}OAuth\\ \end{tabular} & 
% \begin{tabular}[c]{@{}c@{}}Auto.\\ Test\end{tabular} & \begin{tabular}[c]{@{}c@{}}Manual\\ Test\end{tabular} 
% \\
% \hline
% \hline
% Alexa   & \cmark &  \begin{tabular}[c]{@{}c@{}} Public key based\\ (Enforced) \end{tabular} & \cmark & \cmark & \cmark \\
% \hline
% Google & \cmark &  \begin{tabular}[c]{@{}c@{}} Basic HTTP Auth.\\ (Optional) \end{tabular} &\cmark & \xmark & \cmark  \\
% \hline
% \end{tabular}
% }
% \caption{Features of Amazon Alexa and Google Home.}
% \label{tab:comparison}
% \vspace{-0.2in}
% \end{table}

For the other direction, two main authentications are used for the ``servers'' to authenticate the ``clients''.
First, in step \ding{183}, the cloud needs to authenticate the edge device. This can be done because an ``access token'' has been exchanged when the user first sets up the edge device at home. Second, in step \ding{184}, the endpoint also needs to authenticate the cloud. This step helps the endpoint to ensure that the queries are indeed coming from the Alexa/Google cloud instead of outsiders. We call it ``cloud authentication'', which is done in different ways for Google and Alexa. 
\begin{itemize}
\item {\em Amazon Alexa} uses a public-key based method. The cloud signs its request payload with a private key, and the skill endpoints can verify the signature using the cloud's public key when receiving the request. The verification is {\em required}. 

\item {\em Google Home} does not require authentication between the cloud and the endpoints.  
\end{itemize}

\para{Security Vetting Process.}
To make sure the skill and the endpoint are implemented properly, there are two types of vetting deployed by Amazon Alexa and Google Home. 

\begin{itemize}
\item {\em Automated Skill Vetting.} Alexa requires a skill to pass a series of tests before allowing the skill to enter the app store. The test is fully automated and covers both functional tests and security tests. Google Home, however, doesn't have an automated test for the skill endpoint. 

\item {\em Manual Vetting.} For both Alexa and Google, there are dedicated teams that perform manual vetting on the skill before publishing the skill.
\end{itemize}
% Before finally appearing on Alexa and Google market,
% the skill needs to be submitted for Alexa and Google teams to do a manual test.
% However, based on our experience, 
% the manual test is about 
% the logic and the experience of the skill,
% not the security of the endpoint.
% Low adoption rate -> 3 research Questions
% How to answer 3 Questions
% Ethics

\para{Our Focus.} 
Our goal is to perform a case study to learn how effective the security vetting is and how well it helps developers to develop secure skills. We primarily focus on {\em cloud authentication} between the cloud and the third-party endpoints because it is {\em entirely implemented by the skill developers}. In addition, the cloud-to-endpoint communication is also the key reason why smart-assistant skills are fundamentally different from conventional mobile apps --- there is a need for the cloud in the middle to translate a voice command to an API call. One might ask --- why it is Amazon/Google's responsibility to ensure the third-party servers authenticate the cloud. We argue that app developers often lack the security experience~\cite{ieeesp16codesecurity, usenix17, 1334896}. As such, Amazon and Google, as the app store operators, are in the best position to act as the gatekeeper to ensure all the developer components that interact with their infrastructure ({\em i.e.}, the cloud) are securely implemented. This helps to create a more secure ecosystem that is in the best interest of both Amazon/Google and developers/users. Note that other authentication steps that are implemented by Google/Amazon teams (instead of developers) are not considered in this paper (potential future works).

\section{Automated Skill Vetting}
\label{sec:tests} 
We start with Alexa's automated skill vetting (since Google Home does not have an automated skill vetting process). Our goal is to understand what security tests are running against the skill under vetting. Then we build our own skills, deliberately leave mistakes, and examine if the automated tests can detect them.

% Recall that Google does not have automated tests, and thus this analysis is applied to Alexa only.
% According to the Amazon official document, 
% in order to host a skill,
% the endpoint needs to meet a few security requirements~\cite{custom_skill}.
% Alexa sets up automatic functional test to
% make sure skill endpoints meet those requirements.
% Before final release,
% both Alexa and Google have another round of manual test. 
% Given the Alexa functional test and manual tests, 
% we want to know what exactly they are testing.
% We are also curious how strict and representative those tests are.

% Our test methodology has two steps.
% For the step one, 
% we try to reverse engineer the automatic Alexa functional test.
% For the step two,
% we try to release a skill with a vulnerable endpoint
% to see if they can pass the functional test and the manual test.

\subsection{Setting Up Vulnerable Skills}
We implement an Alexa skill with 6 different versions and each version contains different security or functional errors. 

\para{Supported Intents.} Every Alexa skill should support 6 default command-lines defined by Amazon Alexa, and at least 1 custom command-lines defined by the developer. The 6 default command-lines are mapped to 6 built-in {\em intents}. These intents include ``LaunchRequest'', ``StopIntent'', ``CancelIntent'', ``FallbackIntent'', ``HelpIntent'', and ``NavigateHomeIntent'', which are used to perform the basic controls of the skill. We implement the skill to support all 6 default intents and 1 custom intent that takes an integer parameter. 

\para{HTTPS Certificate.}
Both Alexa and Google require HTTPS for the endpoints. Two types of certificates are allowed including standard certificate and wildcard certificate. For our experiment, we test both types of valid certificate, and use a self-signed certificate as the baseline. 

% \begin{enumerate}
% \item {\bf Standard certificate}: a valid certificate from a trusted CA just for the domain of the endpoint. 
% \item {\bf Wildcard certificate}: if the endpoint is hosted under a subdomain ({\em e.g.}, ``{\tt skill.mydomain.com}''), Alexa allows a wildcard certificate (``{\tt *.mydomain.com}'') from a trusted CA.   
% \item {\bf Invalid certificate}: the certificate is not signed by one of Alexa's trusted CAs.   
% \end{enumerate}

% Note that self-signed certificates are not acceptable for skills that need to be published (only works for skills under-development), and thus we don't consider this option. For our experiments, we test the two different types of certificates.

\para{Implementing the Cloud Authentication.} The cloud authentication is used for the endpoints to authenticate the incoming requests from the cloud. 
According to the Alexa documentation, the request from the cloud will contain the signature from the cloud. 
In addition, each request also contains an {\em application-ID} which indicates which application (skill) this request is intended for; and a {\em timestamp}.
Below, we develop 6 different versions of the endpoints: 
\begin{enumerate}
\item {\bf Valid implementation}: For a given request, we validate the cloud signature, 
application-ID, and timestamp before sending a response.  
\item {\bf Ignoring application-ID}: Everything is implemented correctly, except that we ignore the application-ID.  
\item {\bf Ignoring timestamp}: Everything is implemented correctly, except that we ignore the timestamp. 
\item {\bf Accepting all requests}: We do not perform authentication, and always return a legitimate response.
\item {\bf Rejecting all requests}: We drop all the requests.
\item {\bf Offline endpoint}: The endpoint is not online.
\end{enumerate}

% \begin{table}[t]
% \small
% \centering
% \begin{tabular}{|l|l|l|}
% \hline
% Type of Request & Variables & \# Queries\\ 
% \hline \hline
% \multirow{3}{*}{Custom Intent}  & Valid parameter  & 6 \\ 
% & Invalid parameter &   2 \\ 
% & No parameter &   0 \\ \hline

% \multirow{3}{*}{Launch Request$^*$}   & requestIDs  & 11 \\
% &  userIDs &  7 \\ 
% &  deviceIDs &   7 
% \\ \hline

% \multirow{4}{*}{Other Functions}  & HelpIntent &   1\\
% & StopIntent & 1 \\ 
% & CancelIntent & 1 \\ 
% & SessionEnded Request &   1 \\
% \hline

% \multirow{4}{*}{Security Tests} &  Invalid Certificate URL  & 1   \\
% & Invalid Padding &  1 \\
% & Wrong Signature Length &  1 \\ 
% & Empty HTTP Signature &  1 \\ \hline
% \end{tabular}
% \caption{Breakdown of Alexa's testing queries, query variables, and the number of variable values. $^*$For launch requests, the variables are not tested independently, but in certain combinations.}
% \label{tab:requests}
% \vspace{-0.2in}
% \end{table}

% Limin: redid the experiment on 05/05/2019, results stay the same as before.
\begin{table}[t]
\small
\centering
\begin{tabular}{|l|cc|cc|cc|}
\hline
\multirow{3}{*}{Implementation} &  \multicolumn{6}{c|}{Certificate Options} \\ \cline{2-7}
 & \multicolumn{2}{c|}{Standard} & \multicolumn{2}{c|}{Wildcard} & \multicolumn{2}{c|}{Invalid} \\ 

% \cline{2-7}
% & Pass? & \#Req. & Pass? & \#Req. & Pass? & \#Req. \\ 
% \hline \hline
% Valid & \cmark & 28 & \cmark& 28 & \xmark& 0 \\ \hline
% Ignore App-ID &  \cmark& 28 & \cmark& 28 & \xmark& 0 \\ \hline
% Ignore Time & \cmark& 28 & \cmark& 28 & \xmark&0 \\ \hline
% Accept All & \xmark& 28 & \xmark& 28 & \xmark& 0 \\ \hline
% Reject All & \xmark& 38 & \xmark& 38 & \xmark& 0 \\ \hline
% Offline &  \xmark& 0 & \xmark& 0 & \xmark& 0 \\ \hline
\cline{2-7}
& Pass? & \#Req. & Pass? & \#Req. & Pass? & \#Req. \\ 
\hline \hline
Valid & \cmark & 30 & \cmark& 30 & \xmark& 23 \\ \hline
Ignore App-ID &  \cmark& 30 & \cmark& 30 & \xmark& 23 \\ \hline
Ignore Time & \cmark& 30 & \cmark& 30 & \xmark& 23 \\ \hline
Accept All & \xmark& 30 & \xmark& 30 & \xmark& 23 \\ \hline
Reject All & \xmark& 35 & \xmark& 35 & \xmark& 33 \\ \hline
Offline &  \xmark& 0 & \xmark& 0 & \xmark& 0 \\ \hline
\end{tabular}
\caption{Results of Alexa automated test. We have 18 different settings. For each setting, we report whether the skill passed the test, and the number of testing requests that the endpoint received.}
\label{tab:function_test}
\vspace{-0.2in}
\end{table}

\subsection{Skill Testing Results}
We tested our skill with 18 different settings (3 certificates $\times$ 6 endpoint implementations) in September 2019. As shown in Table~\ref{tab:function_test}, standard certificate and the wildcard certificate return the same results. However, when using an invalid certificate (self-signed), even the correct implementation could not pass the test. The test was terminated immediately when the invalid certificate was detected. This result indicates that the automated tests have successfully identified invalid certificates.

As shown in Table~\ref{tab:function_test}, the ``{\em Accepts All}'' implementation failed to pass the test. Analyzing the server logs shows that Alexa cloud has sent a number of queries that carry empty or invalid signatures. 
If we accept these requests, Alexa will determine the endpoint is vulnerable and should not be published in the store. 

However, we notice that the ``{\em Ignore Application-ID}'' and ``{\em Ignore Timestamp}'' implementations both passed the automated test. 
This means that if the endpoint validates the signature but {\em ignores the application-ID or the timestamp}, the skill can still proceed to be published. 
The result raises a major concern. Without validating the application-ID, an endpoint may accept a (malicious) request that is not intended for itself. Our attack in the next section will further exploit this vulnerability.

\section{Spoofing the Cloud}
\label{replay}
The above experiment has two takeaways. First, Alexa enforces the endpoint to validate the signature of the incoming request; This means that published skills only accept incoming requests signed by Alexa. Second, Alexa does not enforce the endpoint to validate the application-ID or the timestamp. This means it's possible a skill endpoint may accept and process outdated requests or requests that are not intended to itself. This can lead to a cloud spoofing attack.

\begin{figure}[t]
\centering
\includegraphics[width=0.4\textwidth]{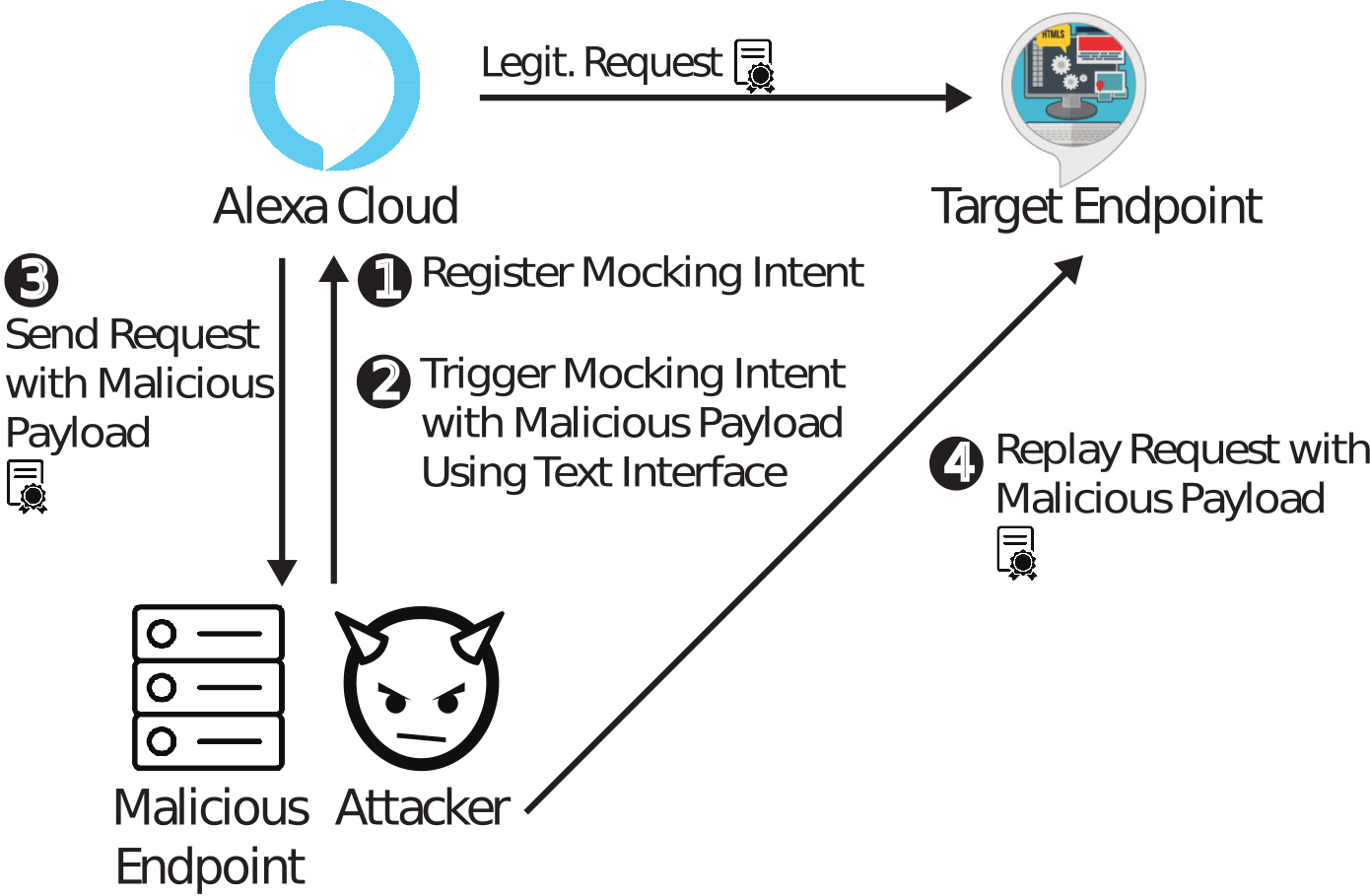}
\caption{The cloud spoofing attack.}
\vspace{-0.12in}
\label{fig:replay}
\end{figure}

\para{Attacking Method.} Given a target skill (the victim), the attacker's goal is to spoof the cloud to interact with the endpoint to insert or retrieve data. We use Figure~\ref{fig:replay} to describe the attack process. The idea is that the attacker builds its own skill, and use this skill to sign the malicious request that will be used for the attack. (\ding{182}) the attacker registers its own skill and the mocking intent. The mocking intent should mimic one of the victim skill's intents (so that the crafted payload is understandable by the victim endpoint). (\ding{183}) Both Alexa and Google have a text interface that allows the developers to type in their command-lines for testing purposes. Using this text interface, the attacker can trigger the cloud to send a request with malicious payload to its own endpoint. (\ding{184}) At this point, the request already carries a valid signature signed by the Alexa cloud. (\ding{185}) The attacker can record this request and then {\em replay} it to the target endpoint. The victim endpoint will believe that the request is from the Alexa cloud. Since the Alexa cloud {\em uses the same private key} to sign all the requests for all the skills, the signature signed for one skill works for other skills too. 

An endpoint can detect this attack if the endpoint checks the {\em application-ID} in the request. Even though the request is signed, the {\em application-ID} inside of the request is still the ID of the attacker's skill. Because the application-ID is inserted by the Alexa cloud before signing the payload, the attacker cannot modify this field.

\para{Proof-of-Concept Experiment.}
The ability to spoof the cloud can lead to different attacks ranging from injecting malicious payload to the victim endpoint to extracting important data from the endpoint. This is different from public-facing web services since skill endpoints are not designed to be public-facing. The servers only expect incoming requests from the cloud. To validate the effectiveness of the spoofing attack, we set up our own target skill {\tt A} as the victim. The victim endpoint {\tt A} is configured to ``ignore application-ID and timestamp''. Then we simulate an attacker by building another skill {\tt B}, and use the endpoint of {\tt B} to collect the malicious requests that will be replayed. We perform this test for both Alexa and Google Home to trigger all 6 default command lines in {\tt A} ({\em e.g.}, to launch or pause the skills). The results show all the attacks were successful.

% \para{Results.}
% We set up the malicious skill and endpoint for both Alexa and Google Assistant,
% record default skill launch request for Alexa and a ``RankIntent'' request for both platforms
% and then forward it to target endpoints.
% All requests are accepted and responded with legitimate responses.
% Figure~\ref{XXX} shows the results.

\para{SQL Injection Attack.} 
Attackers may perform SQL injection attacks on top of the cloud spoofing. The idea is to design a malicious SQL statement and then get the payload signed by the cloud using her own skill. Then the attacker can replay the signed SQL statement to the victim endpoint. To validate the feasibility, we run an attack on {\em our own skill}. We target the skill's Custom Intent that has an integer parameter. The skill server does not have any SQL injection defense ({\em e.g.}, input sanitization). We run a series of SQL injection attacks ({\em e.g.}, inserting data, dropping a table), all of which are executed successfully. In practice, this attack might be more difficult since attackers may not have the full knowledge of the victim endpoints. The attacker needs to guess: (1) the intent name and its parameter name; (2) the name of the target table; (3) the name of the target column. For example, existing SQL injection tools such as SQLMap~\cite{sqlmap} and web vulnerability scanners~\cite{adam-12} would crawl the corresponding websites, find candidate URLs, and issue a large volume of testing queries. It might take hundreds or thousands of automated guesses to search for an injection opportunity (out of the scope of this paper). For Items (2)--(3), there is a way to find related metadata in many mainstream databases. For example, the MySQL database has a metadata table that contains information about all the table names and column names.

\para{Ethical Considerations.} The experiments above are ethical since both the attacker endpoint and the victim endpoint are developed by us. The tests are conducted in the developing mode. There are no other skill endpoints or users involved in the experiments.

\begin{figure}[t]
\includegraphics[width=0.46\textwidth]{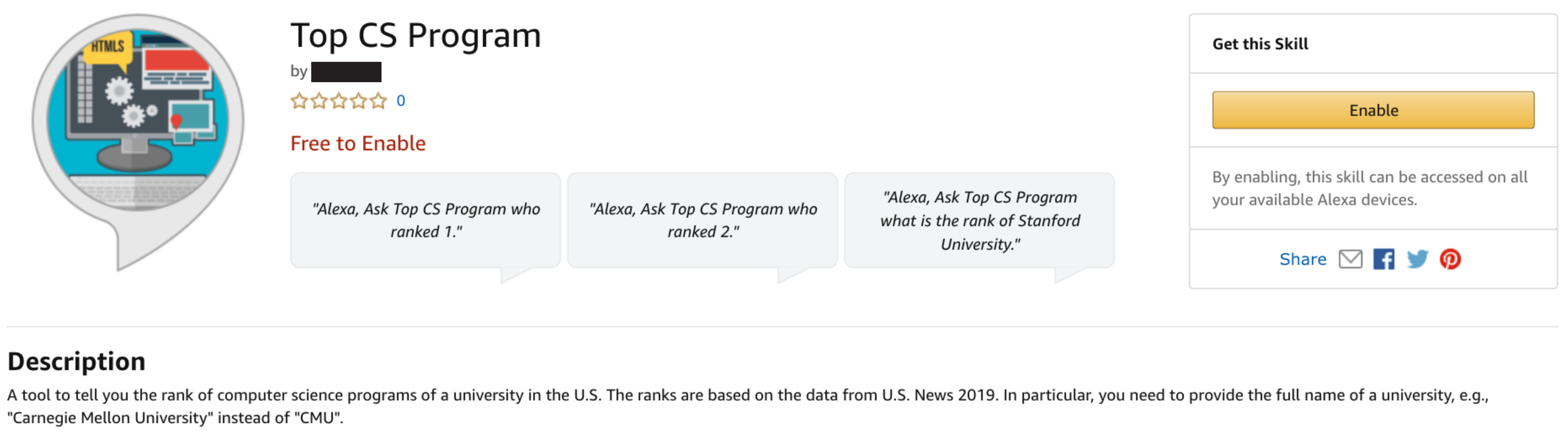}
\vspace{-0.07in}
\caption{Vulnerable skill in the Alexa Skill Store.}
\vspace{-0.1in}
\label{fig:topcsprogram}
\end{figure}

\section{Manual Vetting}
Before publishing, a skill needs to be vetted manually by Amazon and Google teams. To understand the manual vetting process, we send our vulnerable skills (vulnerable to spoofing and SQL injection) for publishing. During the submission process, we did not receive any suggestions related to security issues. Both skills received approval to be released within 1--2 weeks. Figure~\ref{fig:topcsprogram} shows the screenshot of our published skill page (the screenshot for Google is omitted for brevity). 
We immediately took the skill down from both stores after the experiment and informed Google and Amazon about our research. The result suggests that the current vetting process is not rigorous enough to help developers to detect vulnerabilities. 

\para{Ethical Considerations.} We took active steps to ensure research ethics. At the high level, this skill is a ``vulnerable'' skill instead of a ``malicious'' skill. It is supposed to be the victim instead of the attacker in the threat model, which should not introduce any malicious impact. One concern of publishing a vulnerable skill is that the skill may be accidentally used by an innocent user. To avoid this, we have closely monitored our skill endpoint throughout the releasing process. Once the skill received the approval, we immediately performed a quick test to ensure the skill is truly available on the store, and then took it down from the store right away. During this process, we monitored our server and we did not see any incoming requests (except those from our own). This means no real users have ever used this vulnerable skill. In addition, the skill is designed to {\em provide} information (about Computer Science programs in the US), without collecting any user data. Even if a user accidentally used the skill, there is no actual harm.

\section {Alexa Vulnerable Endpoints}
\label{sec:wild}
So far, we show that the security vetting process is not rigorous enough to prevent a vulnerable skill from entering the app store. Next, we focus on {\em Alexa skills} and examine whether there are indeed real-world vulnerable skill endpoints. 
Google Home skills are by default vulnerable to spoofing and thus are omitted in this measurement. 

% In the previous section,
% we prove that the replay attack and the SQL injection attack are practical with skill endpoints.
% However, all of our proof-of-concept experiments are conducted to our own endpoint.
% In this section,
% we examine whether the replay attack vulnerability and the SQL injection vulnerability exist in skill endpoints in the wild.

% back-end server,
% skill endpoints are not visible to users and devices.
% During the development of a skill,
% the developer registers the endpoint to the assistant cloud.
% During normal usage of the skill,
% assistant cloud sends the request on behalf of the device.
% This working process makes locating endpoints challenging.
% Also, we try to look for endpoint information in the audio response
% and introduction web pages for skills.
% However, we didn't find any information about skill endpoints. 

\subsection{Methodology to Locate Skill Endpoints}
For this analysis, we aim to detect endpoints vulnerable to cloud spoofing. We did not further test SQL injection considering the intrusive nature of SQL injection attacks. We face two main challenges. First, the smart assistant devices (edge device) do not directly interact with the skill endpoints. Instead, all the network traffic is first routed to the Amazon cloud. As such, it is difficult for outsiders ({\em i.e.}, researchers) to know the IP or domain name of the endpoint. Second, even if the IP is known, the skill service is not necessarily always hosted under the root path.

\begin{table}[t]
\centering
\small
\begin{tabular}{|c|c|c|c|c|c|}
\hline
% \multirow{3}{*}{System} & 
\multirow{3}{*}{\begin{tabular}[c]{@{}c@{}}IPs enable\\ Port443\end{tabular}} & \multicolumn{3}{c|}{Round 1} & Round 2 & Total \\ 
\cline{2-6}
   & 
   \begin{tabular}[c]{@{}c@{}}Domain \\Set \end{tabular} & \begin{tabular}[c]{@{}c@{}}Candidate\\ Hosts\end{tabular} & \begin{tabular}[c]{@{}c@{}}Vul.\\ EPoints\end{tabular} & \begin{tabular}[c]{@{}c@{}}Vul.\\ EPoints\end{tabular} & \begin{tabular}[c]{@{}c@{}}Vul.\\ EPoints\end{tabular} 
\\ \hline \hline
% Alexa &
\multirow{1}{*}{48,141,053} & 3,196 & 3,346,425 & 122 & 100 & 219
\\ 
\hline
% \cline{1-1} \cline{3-7} 
% Google &  & 639 & 4,155 & 5 & 0 & 5 \\ \hline
\end{tabular}
\caption{Searching results of vulnerable endpoints.}
\label{tab:filter}
\vspace{-0.2in}
\end{table}

% The process to locate skill endpoints.
% We start with IPs that open port 443,
% then we use a IP PTR records dataset to get their hostnames.
% From the term of use and the privacy policay web page,
% we collected 3,196 and 639 second-level domains for Alexa and Google Assitant respectively
% and then use the set to filter possible hosts.
% We successfully lower the possible candidates number to 3,346,425 for Alexa and 4,155 for Google Assistant.
% After scanning,
% we find 174 Alexa skill endpoints,
% however we didn't find any Google Assistant skill endpoint.

\para{Method Overview.} We propose a heuristic-based searching method, based on two intuitions. First, a skill endpoint is required to support HTTPS, which means the port 443 should be open.  Second, an Alexa endpoint should support the default intents such as ``LaunchRequest'' and ``StopIntent''. The response for a default intent request should follow the special JSON format defined by Alexa. As such, we search for vulnerable skill endpoints by scanning the HTTPS hosts with a testing query. The query carries the spoofed ``LaunchRequest'' intent which is a default intent that every Alexa skill should support. We choose this intent because ``LaunchRequest'' won't cause any internal state change or reveal any non-public information.

\para{Implementation.}
Given the large number of HTTPS hosts and the need for {\em guessing the path}, it is not feasible to test a large number of possible paths on all HTTPS hosts. As such, we prioritize search efficiency by sacrificing some coverage. First, we focus on a small set of HTTPS hosts and test many possible {\em paths}. Then we select the {\em most common path} to scan the rest HTTPS hosts. 

For round-1, we select a small set of HTTPS hosts that are more likely to be the skill endpoints. More specifically, we crawled 32,289 Alexa skills pages from Amazon store, and extract their URLs of the privacy Policies. Our hypothesis is that the skill endpoint might share the same {\em domain name} with the privacy policy URL. Note that some skills host their privacy policy on cloud services ({\em e.g.}, ``{\tt amazonaws.com}''). As such, we make a whitelist of web hosting services and only consider the hostname (instead of the domain name) in their privacy policy URLs as the candidate.

Then we test a list of possible paths. We obtain the path information by analyzing the example code on the Developer Forum of Alexa and related question threads in StackOverflow~\cite{stack_alexa, stack_google}. For each host, we test the root path ``/'', and other possible paths including ``/alexa'', ``/echo'', ``/api'', ``/endpoint'', ``/skill'', ``/iot'', ``/voice'',    ``/assistant'',  and ``/amazon''. 

After round-1, we expect to find some real-world skill endpoints. Then, we select the most common non-root {\em path} name. We use this pathname and the root path to test all the HTTPS hosts that have not been tested in round-1. 

\para{Ethical Considerations.} We have taken active steps to ensure research ethics. First, for each host, we only send a handful of queries which has minimal impact on the target host. Second, as detailed below, we re-use the ZMap scanning results~\cite{zmap13} instead of performing our own network-wise scanning to identify HTTPS hosts. The scope is aligned with ZMap port 443 scanning. We respect Internet hosts that don't want to be scanned by ZMap and did not test these hosts. Third, we only test a non-intrusive Intent that does not cause any internal state change of the skill service or reveal any non-public information.

\subsection{Detecting Vulnerable Skill Endpoints}
We start by obtaining a list of 48,141,053 IPv4 addresses with an open 443 port from ZMap's scanning result archive~\cite{zmap13}. 

\para{Round-1 Search.} As shown in Table~\ref{tab:filter}, we obtained the Privacy policy URLs from all the 32,289 skills available in the Alexa U.S. skill store. We extracted 3,196 unique domain names. By matching these domain names with those of the 48 million HTTPS hosts, we got 3,346,425 candidate hosts. 

% This matching process allows us to effectively capture skill endpoints that are hosted under sub-domains ({\em e.g.}, {\tt api.skilldomain.com}). 

By testing the spoofed intent (and candidate paths), we found 122 Alexa skill endpoints that provided a valid response. Here we use an IP address to uniquely represent an endpoint server. In fact, we have identified 174 URLs that have returned a valid response. Some of the URLs are actually mapped to the same IP address.

% \begin{table}[t]
% \centering
% \small
% \begin{tabular}{|l|l|}
% \hline
% Path & \# of Matched URLs \\
% \hline \hline
% /alexa & 88 \\ \hline
% / & 30 \\ \hline
% /echo & 17 \\ \hline
% /api & 14 \\ \hline
% /endpoint & 13 \\ \hline
% /skill & 12 \\ \hline
% /voice & 0 \\ \hline
% /amazon  & 0 \\ \hline
% /assistant,  & 0 \\ \hline
% /iot  & 0 \\ \hline \hline
% Total & 174 \\
% \hline
% \end{tabular}
% \caption{Alexa endpoint path names and the number of endpoint URLs from the Round-1 search.}
% \label{tab:path}
% \end{table}
% Note that one endpoint can support multiple paths and domain names, and thus multiple endpoint domains can be resolved to the same IP address. This is why the total path count is higher than that of the number of endpoints shown in Table~\ref{tab:filter}.

\para{Round-2 Search.} Based on the round-1 result, we find that 
``{\tt /alexa}'' is the most common path (88 out of 174), followed by the root path (30 out of 174). Next, we use these two paths to perform the round-2 searching. As shown in Table~\ref{tab:filter}, we discovered 100 {\em additional} vulnerable endpoints.

% % 1,265,917,039
% However, we only have a list of 48 million IP addresses that open port 443.
% To obtain their hostnames, we use another dataset that provides all IPv4 PTR lookups 
% for all non-blacklisted/private IPv4 addresses.
% It contains the corresponding hostnames for 12.6 billion IPv4 address.

% Note that because some skills host their privacy policy using cloud services like AWS and CloudFront,
% the second-level domain set contains ``amazonaws.com'' and ``cloudfront.net''.
% As a result, all hostnames hosted using these cloud services are included in our possible candidates.
% We succeed in lowering the possible candidates to 3.3 million for Alexa
% and \fixme{4155} for Google Assistant.

\begin{figure}[t]
\centering
\includegraphics[width=0.37\textwidth]{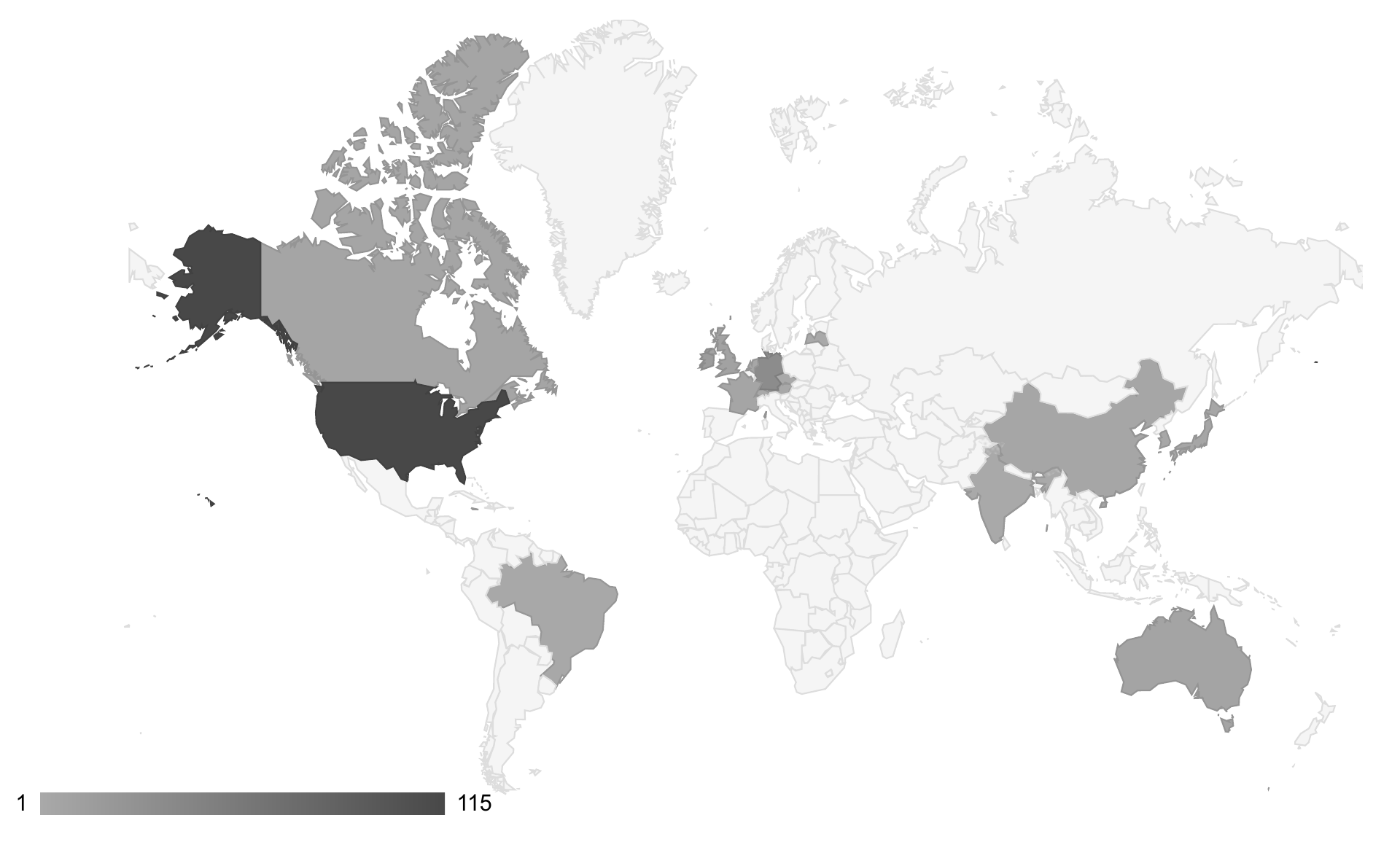}
\caption{Geo-location of 219 vulnerable endpoints.}
\label{fig:map}
\vspace{-0.13in}
\end{figure}

\para{Vulnerable Skill Endpoints.} 
From the two rounds of searching, we detected in total 219 vulnerable Alexa endpoints. It should be noticed that searching result is only a lower-bound considering the incomplete guessing of pathnames. There could be even more vulnerable skill endpoints in the wild.

% \begin{table}[t]
% \small
% \centering
% % \resizebox{0.49\textwidth}{!}{%
% \begin{tabular}{|l|l|l|}
% \hline
%   Country (\#) & Region (\#) & Organization (\#) \\
% \hline
% \hline
%  US (115) & Texas (40) & Amazon (119) \\ 
%  DE (35) & Oregon (26) & Database by Design (22) \\ 
%  IE (16) & New York (23) & DigitalOcean (12) \\ 
%  GB (9) & Hessen (15) & OVH SAS (4) \\ 
%  AU (7) & California (12) & Hetzner Online (4) \\ 
%  CA (6) & New South Wales (7) & Netcup (4) \\ 
%  FR (6) & London (6) & Host Europe (4) \\ 
%  JP (5) & Bayern (6) & Pathway Commun. (4) \\
%  CN (4) & New Jersey (4) & Evocative (2) \\ 
%  KR (3) & Tokyo (3) & Baidu (2) \\ \hline
% \end{tabular} % }
% \caption{The top 10 countries, top 10 regions and top 10 organizations of the 219 vulnerable Alexa endpoints.}
% \label{ip-top}
% \end{table}

Figure~\ref{fig:map} illustrates the geolocation distribution of these vulnerable endpoints based on their countries. 
We observe that more than half of vulnerable endpoints (115, 52.5\%) are located in the United States, followed by Germany (35, 16.0\%) and Ireland (16, 7.3\%). The top 3 countries cover 75.8\% of all vulnerable endpoints. 

% The distribution of endpoint regions is more evened out where Texas has the most vulnerable endpoints. 
% Not too surprisingly, most of the vulnerable endpoints are servers of Amazon Web Services (AWS) set up by the skill developers (119, 54.3\%).

% \begin{table}[t]
% \centering
% \small
% \caption{\hang{The category of vulnerable skills.}}
% \label{skill-cate}
% % \resizebox{\textwidth}{!}{%
% \begin{tabular}{|l|l|}
% \hline
% Category & \# \\
% \hline \hline
% Smart Home & 5 \\ \hline
% Game \& Sports & 2 \\ \hline
% Audio & 1 \\ \hline
% Car Controller & 1 \\ \hline
% Mail & 1 \\ \hline
% Education & 1 \\ \hline
% \end{tabular}%
% % }
% \end{table}

\subsection{Case Studies}
To understand what the vulnerable endpoints represent, we send another spoofed ``HelpIntent'' request to each endpoint. The returned information helps to identify the actual skills. Some vulnerable skills are less ``safety-critical'' which are related to games, sports, and news. However, there are indeed skills that are providing critical services. For example, one vulnerable skill on Alexa is used for controlling electric cars. At least three vulnerable skills are from online banking services. A number of vulnerable skills are used to control other smart-home or IoT devices to turn on/off the bedroom light, adjust the air purifier and thermostats, set an alarm for the home security system, and keep track of water and electricity usage. Leaving their endpoints vulnerable to cloud spoofing poses real threats to users. We give a few specific examples below.

\para{Smart Home.} ``Brunt'' is an automated home furnishing accessory company, and its products include smart plugs, wireless chargers, air purifiers, blind controllers, and power sockets. The vulnerable skill ``Brunt'' supports turning on and off Brunt devices and changing their configurations. 

\para{Connected Cars.} ``My Valet'' is one of the most popular skills that can control Tesla cars. The skill can be used remotely to lock and unlock the car, obtain information of the car's current location, and open the roof and the trunk. Note that the skill is not officially developed by Tesla. To use it, My Valet redirects a user to My Valet's own website and ask for the user's password for her Tesla account (instead of using OAuth). This is already a questionable method to access user account. In addition, the skill's endpoint is vulnerable to cloud spoofing, exposing itself to further risks. 

\para{Social \& Communication.} ``Newton Mail'' is a cross-platform email client, supporting reading recent emails and other common operations such as snoozing and deletion of an email.

\section{related work}

\para{IoT Security \& Privacy.} 
With the wide adoption of IoT devices, security and privacy have become a pressing issue~\cite{zhang2018homonit, naeini2017privacy, chow2017last, fernandes2017security, cheniotfuzzer, kafle2019study, apthorpe2019evaluating, celik2019iotguard, ccs19_zuo}.
A large body of existing works focuses on the security of software and firmware of the IoT devices~\cite{hernandez2014smart, smart_ransomware, jia2018novel}, and measurements of IoT botnets~\cite{ccetin2019cleaning, herwig2019measurement}.
A more related direction looks into the {\em user authentication} schemes of IoT devices~\cite{feng2017continuous, zhang2017hearing, zhang2016voicelive, tian2017smartauth, jia2017contexiot}. Due to a lack of authentication, malicious parties may inject command-lines~\cite{smart_command, bispham2018nonsense} and control the device through inaudible voice~\cite{zhang2017dolphinattack}. 
Our work is complementary by focusing on cloud authentication (instead of user-end authentication). 

% Our work is different in since we look into smart-home assistant systems and examine the interaction between the cloud and third-party endpoints. 

% A recent works that incorrect endpoint side checks can lead to severe security vulnerabilities including password brute-forcing, leaked password probing, and security access token hijacking~\cite{zuoautomatic}. 

\para{Third-party Applications for IoT Devices.} Researchers have analyzed the third-party applications for Samsung-owned SmartThings~\cite{fernandes2016security, fernandes2016flowfence, ding2018safety}. For example, a recent study on 185 SmartThings applications found 37 risky physical interaction chains~\cite{ding2018safety}. Other researchers also examined the attack surface of IoT triggered IFTTT applets~\cite{ccs19_wang, celik2019iotguard}. A study~\cite{bastys2018if} showed that 30\% of the IFTTT applets have violated the privacy expectations. Two recent works studied Alexa and Google Home skills with a focus on their voice interfaces~\cite{kumar2018skill, zhangdangerous}. Our work is different since we focus on authentication between the cloud and third-party endpoints. 

\para{App Developers and Security Coding.} A related body of work focuses on understanding the mistakes made by app developers and how to help developers to improve security. Most existing works focus on Android apps. For example, researchers show that many poorly implemented security mechanisms in mobile apps 
are due to developers who are inexperienced, distracted or overwhelmed~\cite{ieeesp16codesecurity} or copying and pasting code from online forums~\cite{1334896}. Researchers also find developers asking more permission than they need~\cite{Nguyen:2017:, Felt:2011:APD:2046707.2046779, felt2011permission} or failing to use cryptographic API correctly~\cite{chin2011analyzing, egele2013empirical, enck2011study, fahl2012eve, fahl2013rethinking}. Even with tools to help the developers in the Android development environment, it is often not enough to prevent insecure apps~\cite{Nguyen:2017:}. While most existing works are focused on smartphone apps, we for the first time investigate this issue in smart assistant systems. Our result shows that more work needs to be done to help the developers.

\section{Discussion \& Conclusion}
\label{sec:discuss}
The problem described in the paper comes from the insufficient security vetting process before releasing the applications into the market. More specifically, there is a confusion between cloud and application-level authentication in Alexa's automated testing. Alexa enforced an endpoint to verify the cloud identity but did not enforce the verification of the application identity. 
This makes endpoints vulnerable to replayed requests that were intended for other applications ({\em e.g.}, the attacker's skill). We show that developers indeed left this vulnerability in published skills. Part of the reason is that smart-home devices involve complex interactions between different entities, making it more challenging for inexperienced developers to write code securely. To this end, platforms such as Google and Amazon have a bigger responsibility to rigorously test third-party skills before allowing them to enter the stores. 

We have reported our findings to the Alexa and Google Home team and informed them about our experiments. The countermeasure is to implement dedicated skill tests and enforce developers to check the application-ID and the timestamp. We also plan to notify the corresponding skill developers to address the issue. 

\para{Limitations and Future Directions.} Our work has a few limitations. First, our searching only covers a limited number of ``paths''. The number of vulnerable endpoints can only be interpreted as a lower bound. Future works may extend the scanning scope. Second, we only confirmed that the endpoints were vulnerability to cloud spoofing attacks. We did not further test SQL injection attacks for ethical considerations. Future works (with the consent of the skills developers) may explore the feasibility of actual attacks. Third, we only look into the cloud-to-endpoint authentication. Future work can further examine other authentication steps ({\em e.g.}, account linking, OAuth) and other security and privacy aspects ({\em e.g.}, HTTPS implementation, permission management, privacy policies). 

An open question is how to design the security vetting process to effectively help developers. There are two main future directions for exploration. First, we need to improve the coverage of the automated tests to perform more security checks. Second, we need to provide informative and actionable feedback to developers. We could even integrate the checking-and-feedback mechanism into the software development kit (SDK) to improve the security during the skill development process.

\newpage
\bibliographystyle{acm}
\bibliography{hang.bib,wang.bib}

\end{document}